\documentclass[]{aa}
\usepackage{graphics,latexsym}     
\begin{document}
\thesaurus{02(12.07.1, 11.03.1, 12.03.2, 03.13.6)}
\title{Cluster Mass Profiles from Weak Lensing II}
\author{Lindsay J. King$^{1,2}$, Peter Schneider$^{1,2}$}
\institute{1: IAEF der Universit{\"a}t Bonn, Auf Dem H{\"u}gel 71, Bonn,
Germany\\
2: MPA, Karl-Schwarzschild Str 1, Garching bei M{\"u}nchen, Germany}
\date{}
\authorrunning{L.J.King \& P.Schneider}
\maketitle
\begin{abstract}
When a cluster gravitationally lenses faint background galaxies, its 
tidal gravitational field distorts their shapes (shear effect) and its
magnification effect changes the observed number density. In 
Schneider, King \& Erben (2000) we developed likelihood
techniques to compare the constraints on cluster mass profiles 
that can be obtained using the shear and magnification information. 
This work considered circularly symmetric power-law
models for clusters at fairly low redshifts where
the redshift distribution of source galaxies could be neglected. 
Here this treatment is extended
to encompass NFW profiles which are a good description of clusters from
cosmological N-body simulations, and NFW clusters at higher redshifts
where the influence of various scenarios for the knowledge of the 
redshift distribution are examined. Since in reality the overwhelming 
majority of clusters have ellipsoidal rather than spherical profiles, 
the singular isothermal ellipsoid (SIE) is investigated. We also
briefly consider the impact of substructure on such a likelihood analysis.

In general, we find that the shear information provides a better
constraint on the NFW profile under consideration, so this becomes the
focus of what follows. The ability to 
differentiate between the NFW and power-law profiles strongly depends
on the size of the data field, and on the number density of galaxies
for which an ellipticity can be measured. Combining Monte Carlo 
simulations with likelihood techniques is a very suitable way to 
predict whether profiles will be distinguishable, given the
field of view and depth of the observations. 

For higher redshift NFW profiles, there is very little reduction
$(\sim1.5\%)$ in the dispersion of parameter estimates when
spectroscopic redshifts, as opposed to photometric redshift estimates, 
are available for the galaxies used in the lensing analysis. 


\end{abstract}

\def\A{{\cal A}}
\def\eck#1{\left\lbrack #1 \right\rbrack}
\def\eckk#1{\bigl[ #1 \bigr]}
\def\rund#1{\left( #1 \right)}
\def\abs#1{\left\vert #1 \right\vert}
\def\wave#1{\left\lbrace #1 \right\rbrace}
\def\ave#1{\left\langle #1 \right\rangle}
\def\arcsecf {\hbox{$.\!\!^{\prime\prime}$}}
\def\arcminf {\hbox{$.\!\!^{\prime}$}}
\def\bet#1{\left\vert #1 \right\vert}
\def\vp{\varphi}
\def\vt{{\vartheta}}
\def\d{{\rm d}}
\def\eps{{\epsilon}}
\def\vc{\vec} 
\def\s{{\rm d}}
\def\s{{\rm s}}
\def\t{{\rm t}}
\def\E{{\rm E}}
\def\L{{\cal L}}
\def\q{{\rm \i}}

{\catcode`\@=11
\gdef\SchlangeUnter#1#2{\lower2pt\vbox{\baselineskip 0pt \lineskip0pt
  \ialign{$\m@th#1\hfil##\hfil$\crcr#2\crcr\sim\crcr}}}
}
\def\gtrsim{\mathrel{\mathpalette\SchlangeUnter>}}
\def\lesssim{\mathrel{\mathpalette\SchlangeUnter<}}      

\section{Introduction}

Gravitational lensing provides an invaluable means to investigate how
luminous and dark matter is distributed in the Universe, from inside
our own Galaxy to cosmological scales [e.g. Alcock (2000), 
Keeton et al. (1998), van Waerbeke et al. (2000)]. In this paper we focus on clusters in the weak lensing regime, where the number density and shapes of faint 
background galaxies are changed through the magnification and shear
effects respectively. These signatures can be used to determine the 
projected mass distribution of the lensing cluster.

The pioneering work of Kaiser \& Squires (1993) describes how to
obtain a parameter-free reconstruction of a mass distribution, and 
this technique has been applied to many clusters 
[e.g. Fischer \& Tyson 1997; Clowe et al. 1998; Hoekstra et al. 1998; Clowe et al. 2000; 
Hoekstra, Franx \& Kuijken 2000]. In contrast to parameterised models, the
interpretation of such mass maps is difficult since
the error properties of non-parametric models are poorly understood.
Although deriving 
parameterised cluster models from weak lensing data may not reveal 
the lensing mass distribution in the same detail, it enables 
different families of models to be explored, and statistical comparisons 
between clusters to be made.

In Schneider, King \& Erben (2000; hereafter SKE) we investigated 
parameterised cluster models and developed likelihood techniques to 
quantify the accuracy with which parameters can be recovered using 
the magnification and shear information. Two simple but 
generic families of power-law models for the surface mass density were 
studied. The basic model has a monotonically decreasing form, and is 
defined outside the Einstein radius ($\theta_{\E}$), 
which marks the transition between the weak and strong lensing regimes. 

Here we extend this work to encompass two other parameterised 
lens models that are commonly used to describe cluster mass profiles: 
the NFW profile (Navarro, Frenk \& White 1996;1997), and the singular
isothermal ellipsoid (SIE) as discussed by Kormann, Schneider \&
Bartelmann (1994). The impetus for considering the NFW profile is 
that it is a good description of the radial density profiles of 
virialised dark matter halos formed in cosmological simulations of 
hierarchical clustering. The SIE model allows us to investigate 
non-spherically symmetric dark matter distributions.
The influence of uncertainty in the redshift distribution of 
galaxies used in the lensing analysis is examined, in the 
context of NFW clusters at higher redshifts. 

A question of fundamental importance to the 
study of galaxy and cluster formation, and to the nature of dark
matter itself, is how profiles can best be described [e.g. Moore et al.
2000]. Therefore, we investigate whether we can distinguish between 
NFW and power-law models, and the constraints that can be placed on
SIE models.

We also briefly address the influence of substructure on our analysis, 
by comparing the smooth NFW profile with a toy model containing substructure.

The structure of this paper is as follows: the notation and lensing 
relationships used are outlined in Section 2, followed by the basis of 
using the magnification and shear methods in constraining a cluster
mass profile. In Section 3, the likelihood 
functions developed in SKE are briefly
presented. Moving to clusters at higher redshifts, several scenarios
for our knowledge of the source redshift distribution are described.
Monte Carlo simulations were performed in order to check the validity of the
analytic treatment, and to deal with cases where the redshift
distribution is important; a prescription for these is given in
Section 4. The relevant properties of the lens models considered in this work
are also given in this Section. 

The results from the likelihood analysis in the case of lower 
redshift clusters are presented in Section 5.
For the NFW profile, proceeding as in SKE, we check the validity of 
the assumption of $\chi^{2}$ statistics. Also, a brief 
comparison is made between the 
accuracy of parameters recovered using the magnification and shear 
information. Then the shear method becomes the focus and
we ask whether it is possible to distinguish between the NFW and 
other profiles, using our likelihood analysis. 
Moving to the SIE model, we want to ascertain how well the axial ratio
of a mass distribution is recovered -- can we detect deviation from
circular symmetry and put constraints on the position angle of the
lens model? The results for NFW clusters at
higher redshifts, for different degrees of source 
redshift information, are presented in Section 6. Lastly, we consider
what effect adding substructure to the smooth NFW model has
on our analysis. In the final Section we
summarise our conclusions.

\section{Shear and magnification methods}

In this Section the notation used throughout this paper is introduced,
along with the basic definitions and relationships. Then the
methods for constraining the mass profile of a lens 
using the shape distortions and the change in the number density of
background galaxies are outlined.

\subsection{Notation, basic definitions and relationships}
Throughout, standard lensing notation is used [eg Schneider, Ehlers \&
Falco 1992; Bartelmann \& Schneider (2000)].

The surface mass density of a lens at position $\vc\theta$  
is denoted by $\Sigma (\vc\theta)$ and the critical surface mass density
of a lens at redshift $z_{\rm d}$ for sources at redshift $z$ by 
$\Sigma_{\rm crit} (z)\equiv \frac{c^{2}}{4\pi
G}\frac{D_{\s}}{D_{\d}D_{\d\s}}$, where $D_{\s}, D_{\d}, D_{\d\s}$ are 
the observer--source, observer--lens and lens--source 
angular diameter distances respectively. The 
dimensionless surface mass density of a lens, $\kappa(\vc\theta,z)$, is 
the ratio of $\Sigma / \Sigma_{\rm crit}$, and a Poisson-like equation
relates the deflection potential
$\psi(\vc\theta)$ to $\kappa$ 
\begin{equation}
\nabla^2\psi=2\kappa\; .
\end{equation}
The complex shear is a combination of second derivatives of the
potential 
\begin{equation}
\gamma=\gamma_1+{\rm i}\gamma_2 =(\psi_{11}-\psi_{22})/2+{\rm i}\psi_{12}
\end{equation}
 (the subscript indices denote partial derivatives 
with respect to the position $\vc\theta$ on the sky). Further, 
$g=\gamma/(1-\kappa)$ is the complex reduced shear. The magnification
of an image is the inverse of the Jacobian determinant of the lens
equation, 
\begin{equation}
\mu(\vc\theta)=[\det \A(\vc\theta)]^{-1};~~~~
\det\A=(1-\kappa)^2-|\gamma|^2\; .
\end{equation}  
In Section 4 the expressions specific to the lens model families
considered here are given. 

The strength of a lens depends on the relative redshifts of the
observer, lens and source, and as in 
Seitz \& Schneider (1997) it is convenient to introduce 
a redshift dependent lensing strength factor $w(z)$. Then we can 
express $\kappa(z)=w(z)\kappa_{\infty}$ and $\gamma(z)=w(z)\gamma_{\infty}$,
where $\kappa_{\infty}$ and $\gamma_{\infty}$ correspond to
quantities at position $\vc\theta$ for hypothetical sources located at
$z=\infty$. In the Einstein-de Sitter $\Omega=1, \Lambda=0$ cosmology
which we adopt,
\begin{equation}
w(z)=\frac{\sqrt{1+z}-\sqrt{1+z_{\d}}}{\sqrt{1+z}-1}~;~~z>z_{\d}
\end{equation}
-- of course, $w(z)=0$ when $z<z_{\d}$ since the source is not lensed. 
It follows that $g$ and $\mu$ can be written as
\begin{equation}
g(\theta,z)=\frac{w(z)\gamma_{\infty}}{1-w(z)\kappa_{\infty}}
\end{equation}
and
\begin{equation}
\mu(\theta,z)=\frac{1}{[1-w(z)\kappa_{\infty}]^{2}-[w(z)\gamma_{\infty}]^{2}}
\end{equation}
respectively. In SKE we considered the case of fairly low 
redshift clusters where $w(z)$ is nearly constant for most galaxies
used in the lensing analysis, and the differential probability
distribution $p(w){\d}w$ (equivalent to $p(z){\d}z$) is a very
narrow distribution. 
Then, the redshift distribution of the source 
galaxy population can safely be neglected, approximating them to be
located at a redshift corresponding to the mean value of $w(z)$. 
If the cluster is at a higher redshift, $\gtrsim 0.25$ say, then the redshift
distribution of the galaxies becomes important and this
sheet approximation is no longer robust (Bartelmann \& Schneider 2000). 

Throughout, we take $H_{0}=65~{\rm km~s}^{-1}{\rm Mpc}^{-1}$. 
 
\subsection{The basis of the magnification and shear methods}

\subsubsection{Shear method}

The galaxy ellipticity $\eps$ is defined by a complex number whose
modulus is $|\eps|=(1-r)/(1+r)$, in the case of elliptical isophotes
with axis ratio $r\le 1$, and whose phase is twice the position angle of 
the major axis. A Gaussian probability distribution with
dispersion $\sigma_{\eps^{s}}$ is adopted for the ellipticity

\begin{equation}
p_{\epsilon^{\s}}=
\frac{\exp\rund{-|\eps^{\s}|^{2}/\sigma_{\eps^{\s}}^{2}}}
{\pi\sigma^{2}_{\eps^{\s}}\left[1-\exp\rund{-1/\sigma_{\eps^{\s}}^2}\right]}\;.
\end{equation}

A transformation relates the source ($\eps^{\s}$) and image ($\eps$) 
ellipticities, which are changed by the tidal 
gravitational field of the lens. We focus on the non-critical regime
(det $\A>0$) where

\begin{equation}
\eps={\eps^{\s} + g\over 1+g^{*}\eps^{\s}}\; .
\label{eps}
\end{equation}
%
%

The lensed and unlensed probability distributions are related through
\begin{equation}
p_{\epsilon}=p_{\eps^{\s}}\left
|\frac{{\rm d^{2}\epsilon^{\s}}}{{\rm d^{2}\eps}}\right |\;,
\label{ninfo}
\end{equation}
when the distribution in source redshift is unimportant [see
Eq.(\ref{redinfo}) for the more general form]. 

It can be shown that the expectation value for the lensed ellipticity 
$\ave{\eps}=g$ in the non-critical regime (e.g. Schramm \& Kayser
1995), and that $\ave{\eps}=1/g^{*}$ in the critical regime (Seitz \&
Schneider 1997). This is the basis
of the using the distorted images of background galaxies to constrain
the cluster model. 

\subsubsection{Magnification method}

The magnification method is based on the change in the local number
counts of background galaxies by the magnification bias (e.g. Canizares 1982). The local cumulative number counts
$n(\vc\theta;S)$ above flux limit $S$ are related to the unlensed counts
$n_{0}(S)$ by
\begin{equation}
n(\vc\theta;S)={1\over \mu}n_{0}\,\rund{S\over\mu}\;.
\end{equation}
If we assume that the number counts locally follow a power law of      
the form $n_0\propto S^{-\beta}$, then
\begin{equation}
n(\vc\theta)=n_{0}\,\mu(\vc\theta)^{\beta-1}
\end{equation}
at any fixed flux threshold. This implies that if the intrinsic counts are
flatter than $1$, then the lensed counts will be reduced relative to
the unlensed ones. 

\section{Likelihood treatment}

The goal of our likelihood treatment is to obtain the best-fitting
parameters and their error estimates for simulated or real observations 
[see for example Press et al. (1992)]. Generally, it is more sensible 
computationally to minimise log-likelihood functions to obtain the
best-fitting parameters, since a large 
number of galaxies is involved. Whereas a log-likelihood function 
pertains to a single data set or realisation, ensemble averaged log-likelihood
functions represent the average over many such realisations, and 
enable the characteristic (expected) errors associated with the
recovered parameters to be estimated. 

We consider cluster lenses described by parameterised models, with 
true parameters $\pi_{\t}$, best-fit parameters $\pi_{\rm max}$, 
and trial parameters (en route to minimisation) $\pi$ throughout. 
Quantities that refer to the true
model value are subscripted with a ``t", and those referring to
observed values are subscripted with an ``$i$". We assume that in an
aperture centred on 
the cluster there are $N_{\mu}$
images of background galaxies at positions $\vc\theta_{i}$, above a
given flux limit, and $N_{\gamma}$ images at positions
$\vc\vt_{i}$ for which an ellipticity can be measured. There can be an
overlap between the two sets of galaxies.

\subsection{Likelihood functions}
Here the expressions for the likelihood functions and
ensemble averaged log-likelihood functions are reproduced; 
the reader is referred to SKE for the details and derivations.
Minimising these functions gives $\pi_{\rm max}$, the
most likely parameters given the observations.
In the absence of a redshift distribution for the background sources, 
the magnification log-likelihood function is
\begin{equation}
\ell_{\mu} = n_{\mu}\int{\rm d}^{2}\theta[\mu(\vc\theta)]^{\beta-1}+(1-\beta)\sum_{i=1}^{N_{\mu}}{\rm\ln\;}\mu(\vc\theta_{i})\;.
\label{likmag}
\end{equation}
The noise in the magnification method is due to Poisson 
noise on the number of galaxies in the aperture. The first term in
Eq.(\ref{likmag}) dominates the expression, and gives the number of
galaxies expected in the aperture, given $\pi$. 
Given this number of galaxies, the second term depends on how they are 
distributed. One noteworthy caveat of the magnification method that we
uncovered in SKE is that it requires accurate knowledge of the
unlensed number density, $n_{\mu}$.

The shear log-likelihood function is
\begin{equation}
\ell_{\gamma}=-\sum_{i=1}^{N_{\gamma}}{\rm\ln\;}p_{\epsilon}(\epsilon_{i}|g(\vc\vt_{i}))\;,
\label{liksh}
\end{equation}                       
and the noise in this method arises from the intrinsic dispersion in the galaxy
ellipticity distribution, $\sigma_{\eps^{\s}}$.  
Eq.(\ref{liksh}) just depends on how probable particular lensed
ellipticities are, given the $\pi$ under consideration. Unlike the
magnification method, the shear method does not require that the
unlensed number density is accurately known. 

Eq.(\ref{liksh}) can be used numerically with the exact probability
distribution for $p_{\eps}$, but for
analytic work the log-likelihood function $\ell_{\gamma}$ can be dealt
with as
\begin{equation}
\ell_{\gamma}=\sum_{i=1}^{N_{\gamma}}\frac{(\eps_{i}-g(\vc\vt_{i}))^{2}}{\sigma_{\eps}^{2}(g(\vc\vt_{i}))}+2\ln\sigma_{\eps}(g(\vc\vt_{i}))\;,
\label{shan}
\end{equation}
where $\sigma_{\eps}\approx(1-|g|^{2})\sigma_{\eps^{s}}$.
For $|g|\leq 1.0$, this approximation is accurate to $\approx 5\%$ when 
$\sigma_{\eps}=0.2$. 

The combined shear and magnification log-likelihood is obtained by
adding the respective log-likelihoods:
\begin{equation}
\ell_{\rm tot}=\ell_{\mu}+\ell_{\gamma}\; .
\end{equation} 

As mentioned above, the ensemble averaged log-likelihood functions
allow the characteristic errors for parameters to be derived. 
The ensemble averaged log-likelihood function for the magnification is
\begin{eqnarray}
\ave{\ell_{\mu}}&=&n_{\mu}\int\d^{2}\theta\;[\mu(\vc\theta)]^{\beta-1}\nonumber\\
&+&
n_{\mu}(1-\beta)\int\d^{2}\theta\; [\mu_{\t}(\vc\theta)]^{\beta-1}\,\ln\mu(\vc\theta)\;,
\end{eqnarray}
where $\mu_{\t}(\vc\theta)$ is the magnification determined for $\pi_{\t}$.

For the shear we have
\begin{eqnarray}
\ave{\ell_{\gamma}}&=&n_{\gamma}\int\d^{2}\vt\;[\mu_{\t}(\vc\vt)]^{\beta-1}\nonumber\\
&\times&
\rund{{|g(\vc\vt)-g_{\t}(\vc\vt)|^2+\sigma_{\eps,\t}^{2}(\vc\vt)\over
\sigma_{\eps}^{2}(\vc\vt)}+2\ln\sigma_{\eps}(\vc\vt)}\;,
\end{eqnarray}      
where again $\mu_{\t}, g_{\t}$ and $\sigma_{\eps}(g_{\t})$ are determined
for $\pi_{\t}$. 

We have numerically implemented the expressions for the ensemble
averaged log-likelihood functions which means that for specified 
observing conditions (i.e. $n_{\gamma}$, $n_{\mu}$ and size of the
data field) the characteristic errors on parameterised models 
can easily be obtained. This is helpful when assessing the 
benefits of current telescopes with different cameras, or future 
ground- or space-based instruments.

In Section 5 we demonstrate that the distribution of 
$2\Delta\ave\ell \equiv 2(\ave\ell_{\rm max}-\ave\ell_{\t})$ 
for $\pi_{\rm max}$ tends to a $\chi_{M}^{2}$
distribution, where $M$ is the number of model parameters, so that
error interpretation in the framework of Gaussian distributed errors 
is a good approximation.

\subsection{Including a redshift distribution}\label{redinc}

For clusters at intermediate and high redshifts there are
several ways to proceed when obtaining best-fit parameters, 
depending on our knowledge of the individual 
background galaxy redshifts. In this Section a few scenarios for the
information on the redshift distribution are briefly outlined and in
Section 6 a comparison is made between the dispersion in parameters
recovered under each assumption. 

Let the true redshifts of the background sources, denoted by $z_{t}$,
be drawn from a probability distribution $p(z){\d}z$. Here, the redshift 
probability 
distribution used to generate catalogues of lensed galaxies is taken
from Brainerd et al. (1996):
\begin{equation}
p(z){\rm d}z=\frac{\eta
z^{2}e^{-(z/z_{0})^\eta}}{\Gamma(\frac{3}{\eta})z_{0}^{3}}{\rm d}z\;, 
\label{redshift}
\end{equation}
where below it is assumed that $z_{0}=1/3$ and $\eta=1.0$, resulting 
in $\ave{z}=1.0$.

In its most general form, in the presence of a redshift distribution
$p_{\epsilon}$ becomes
\begin{eqnarray}
p_{\epsilon}(\epsilon|\kappa_{\infty},\gamma_{\infty})&=&\int_{0}^{\infty}p_{\epsilon^{s}}\left(\epsilon^{s}\left[\eps,g(z)\right]\right)\left
|\frac{{\rm d^{2}\epsilon^{s}}}{{\rm d^{2}\epsilon}}\right
|(\eps, g(z))\nonumber\\
&\times&
p_{z}(z){\rm d}z
\label{redinfo}
\end{eqnarray}
and this is inserted into the likelihood function. 
Eq.(\ref{redinfo}) is very difficult to deal with analytically, as
discussed by Geiger \& Schneider (1998) who give some approximations
for the integral. 

In this work we explore a few different routes:
\begin{itemize} 
\item{For the purposes of 
comparison, the ideal case is where {\bf (i) all redshifts are known
exactly}.} 
\item{Second, we consider the situation
where {\bf (ii) photometric redshift estimates} $z_{\rm ph}$ {\bf are available for individual galaxies}.} 
\item{A common practice in mass reconstruction is to assume that 
{\bf (iii) the galaxies used in the lensing analysis are at a single redshift 
$z_{\rm sheet}$}, determined by the arithmetic mean of the weighting
factor $\ave{w}$.}
\item{Another computationally reasonable possibility is to assume that
{\bf (iv) the form of} $p(z){\rm d}z$ {\bf is known}, but not the redshifts of 
individual sources, and to integrate (\ref{redinfo}) numerically.} 
\end{itemize}

The availability of photometric redshift estimates is becoming more 
observationally realistic [e.g. Connolly et al. 1995; 
Fern\'{a}ndez-Soto, Lanzetta \& Yahil 1999; Ben\'{\i}tez 2000] and has
a great impact on lensing (e.g. Dye et al. (2000)). 
Obviously, the smaller
the dispersion $\sigma_{z}$ in $z_{\rm ph}$, the smaller the 
dispersion in $\pi_{\rm max}$. This can be quantified with a simple estimate: 
The lensed probability distribution $p_{\eps}(\eps)$ is convolved with the 
Gaussian describing the distribution of $z_{\rm ph}$ around $z_{t}$.
In the weak lensing limit, 
$g\sim w\gamma_{\infty}$ so it is convenient to work with $w$ rather
than $z$, where the dispersion in $w_{\rm ph}$ is $\sigma_{w}$. 
Further using  $\sigma_{\eps}\sim\sigma_{\eps^{\s}}$ gives
\begin{eqnarray}
p_{\epsilon}(\epsilon)&=&\int_{0}^{1}\frac{{\rm exp}\left(-\frac{|\epsilon -
{w_{\rm
ph}}\gamma_{\infty}|^{2}}{\sigma_{\epsilon^{\s}}^{2}}\right)}{\pi\sigma_{\eps^{\s}}^{2}}\nonumber\\
&\times&
\frac{{\rm exp}\left(-\frac{({w_{\rm
ph}}-{w}_{t})^{2}}{2\sigma_{w}^{2}}\right)}{\sqrt{2\pi}\sigma_{w}}{\rm d}w_{\rm ph}\;.
\end{eqnarray}
The integration can approximately be replaced by one between $-\infty$
and $\infty$ 
yielding 
\begin{equation}
p_{\epsilon}(\epsilon)=\frac{1}{\pi\sigma_{\eps^{\s}}\sqrt{\sigma_{\eps^{\s}}^{2}+2\gamma_{\infty}^{2}\sigma_{w}^{2}}}{\rm exp}\left(-\frac{|\epsilon - {w_{t}}\gamma_{\infty}|^{2}}{\sigma_{\epsilon}^{2}+2\gamma_{\infty}^{2}\sigma_{w}^{2}}\right)\;,
\end{equation}
where the dispersion $\sigma_{w}\approx\sigma_{\rm z}\frac{{\rm d}w}{{\rm
d}z}$. 
Note that the quantity
$\sqrt{\sigma_{\eps}^{2}+2\gamma_{\infty}^{2}\sigma_{w}^{2}}$ can
be considered as an effective dispersion $\sigma_{\rm eff}$. In terms of
the effect on lensing, the ratio of $\sigma_{\rm eff}^{2}:\sigma_{\eps}^{2}$
is of interest. Even for $\gamma_{\infty}$=0.3, and taking a very
large value of $\sigma_{w}$ =0.18, $\sigma_{\rm eff}^{2}$ is only 
15\% larger than for the case when the source redshifts are known exactly.


When sources are assumed to be at $z_{\rm sheet}$, this
results in an expected reduced shear 
$\ave{g_{\rm sheet}}$. The lensed ellipticity probability is
\begin{equation}
p_{\epsilon}(\epsilon|\kappa_{\infty},\gamma_{\infty})=p_{\eps^{\s}}\left(\eps^{\s}\left[\eps,
g(z_{\rm
sheet})\right]\right)\left
|\frac{{\rm d^{2}\eps^{\s}}}{{\rm d^{2}\eps}}\right
|\left(\eps, g(z_{\rm sheet})\right)\;.
\label{sheetinfo}
\end{equation}
It has 
been shown by Seitz \& Schneider (1997) that this assumption leads to a 
discrepancy between the true shear, $\ave{g_{\t}}$, and 
$\ave{g_{\rm sheet}}$ of
\begin{equation}
\frac{\ave{g_{\t}}}{\ave{g_{\rm sheet}}}\approx 1+\left(\frac{\ave{w^{2}}}{\ave{w}^{2}}-1\right)\kappa\;.
\end{equation}

In Section 6 we use Monte Carlo simulations to compare how each of
these assumptions affects the accuracy with which parameterised models
can be fit to lensing data.

\section{Lens models and simulations}

Below we describe how the lensing simulations were carried out. In the
case of lower redshift clusters, 
these simulations were used to check the validity of the 
ensemble average analytic treatment in putting confidence limits on
$\pi_{\rm max}$. For higher redshift clusters, simulations were used in
conjunction with analytic approximations in order to determine the 
accuracy with which parameters can be recovered. Following this, the 
main features of the lens models used in the likelihood analysis are 
outlined.

\subsection{Simulations}

It is assumed that the observations are made in a circular aperture of inner
radius $\theta_{\rm in}$ and outer radius $\theta_{\rm out}$ centred on
the cluster. The number density
of background galaxies for which an ellipticity can be measured, and
can therefore be used for the shear method, is $n_{\gamma}$, and the 
number density that can be used for the magnification method is
$n_{\mu}$. Unless otherwise stated, $n_{\gamma}=30~{\rm arcmin}^{-2}$
(the typical number density used for shear analysis of deep 
ground-based data) and $n_{\mu}=120~{\rm arcmin}^{-2}$ (similar to
Fort et al. 1997).

The expected number of galaxies in the 
aperture $\ave N$ is determined, and a random deviate drawn from a Poisson
distribution of mean $\ave N$ gives the number of galaxies in the 
unlensed galaxy catalogue ${\ave N}_{P}$. These galaxies are 
randomly distributed and individual galaxy ellipticities are drawn 
from a Gaussian probability
distribution with 2-D dispersion $\sigma_{\epsilon^{\s}}=0.2$. When the
background galaxies are assigned redshifts, these are drawn at random
from the distribution of Eq.(\ref{redshift}). At this stage, the
catalogue of unlensed
galaxies contains random $\vc\theta_{i}$, $\epsilon^{\s}_{i}$ and
$z_{t}$ (if applicable). For a given lens model family ${\cal M}$,
with parameters $\pi_{\t}$, the lensed ellipticities for each galaxy 
$\epsilon_{i}$ are obtained using the relationship (\ref{eps}), 
where the expressions for $g_{\cal M}(\pi_{\t},\vc\theta_{i},z_{t})$
are given below for particular lens models.
To account for magnification, a fraction of the sources 
is rejected: if a uniform random deviate [0,1] is larger than 
$\left [\mu_{\cal M}(\pi_{\t},\vc\theta_{i},z_{t})\right
]^{\beta-1}$, the galaxy is excluded
from the catalogue. At this stage, uncertainty in the redshift 
distribution of galaxies is incorporated for the scenarios 
outlined in Section\ref{redinc}. Finally, the catalogue contains
$\vc\theta_{i}$, $\epsilon_{i}$ and $z_{i}$ for each lensed galaxy,
representing a single data set. The log-likelihood functions are 
then minimised to obtain $\pi_{\rm max}$ for the
catalogues, either using the lens model family used to generate the data set, or for a different family of models when we want to see how well families
can be distinguished.

\subsection{The NFW profile}

We can parameterise this profile with a virial radius $r_{200}$, and a 
dimensionless concentration parameter $c$, which are related through a
scale radius 
$r_{\rm s}=r_{200}/c$. Inside $r_{200}$, the mass density of the
halo equals 200$\rho_{\rm c}$, where $\rho_{\rm c}=\frac{3H^{2}(z)}{8\pi G}$ is the critical density of the Universe at the redshift of the halo. The 
characteristic overdensity of the halo, $\delta_{\rm c}$, is related to 
$c$ through
\begin{equation}
\delta_{\rm c} = \frac{200}{3}\frac{c^{3}}{{\rm ln}(1+c)+c/(1+c)}\;.
\end{equation}
Then the density profile is
\begin{equation}
\rho(r) = \frac{\delta_{\rm c}\rho_{\rm c}}{(r/r_{\rm s})(1+r/r_{\rm s})^{2}}\;,
\end{equation}
which is shallower than isothermal ($r^{-2}$) near the halo center and 
steeper than isothermal for $r\gtrsim r_{\rm s}$. 

In general, the assumption that the cluster is in equilibrium becomes 
less valid as its redshift increases. Recently, Jing (2000) 
quantified how well the NFW profile describes both equilibrium and 
{\em non-equilibrium} halos, finding that the profile 
is a good fit to about 70\% of all halos, with the deviation increasing
as the amount of substructure increases. It is interesting to note that
simulated halos with more substructure also require a lower value of 
$c$, which is in agreement with the fits to observations of high
redshift clusters made by Clowe et al. (2000). 

The properties of the NFW profile in the context of gravitational
lensing have been discussed by authors including Bartelmann (1996) 
and Wright \& Brainerd (2000). The radial dependence of the
dimensionless surface mass density as a function of a dimensionless
radial coordinate $x:=r/r_{\rm s}$ is given by:
\begin{equation}
\kappa(x)=\kappa_{k}f(x)
\end{equation}
where
\begin{eqnarray}
f(x<1)&=&\frac{1}{x^{2}-1}\left(1-\frac{2~{{\rm atanh}\sqrt{\frac{1-x}{1+x}}}}{\sqrt{1-x^{2}}}\right)\nonumber\\
f(x=1)&=&\frac{1}{3}\\
f(x>1)&=&\frac{1}{x^{2}-1}\left(1-\frac{2~{\rm atan}\sqrt{\frac{x-1}{1+x}}}{\sqrt{x^{2}-1}}\right)\nonumber
\end{eqnarray}
and
\begin{equation}
\kappa_{k}=\frac{2r_{s}\delta_{c}\rho_{c}}{\Sigma_{c}}\;.
\end{equation}
The mean dimensionless surface mass density inside radius $x$ is:
\begin{equation}
\bar\kappa(x)=\kappa_{k}h(x)\;,
\end{equation}
where
\begin{eqnarray}
h(x<1)&=&\frac{2}{x^{2}} \left(\frac{2~{\rm
atanh}\sqrt{\frac{1-x}{1+x}}}{\sqrt{1-x^{2}}}+\ln\left(\frac{x}{2}\right )\right )
\nonumber\\  
h(x=1)&=&2+2\ln\left(\frac{1}{2}\right)\\  
h(x>1)&=&\frac{2}{x^{2}}\left(\frac{2~{\rm atan}\sqrt{\frac{x-1}{1+x}}}{\sqrt{x^{2}-1}}+{\rm ln}\left(\frac{x}{2}\right)\right)\nonumber   
\end{eqnarray}
The shear $\gamma =\bar\kappa-\kappa$ at position $x$ is:
\begin{equation}
\gamma(x)=\kappa_{k}j(x)\;,
\end{equation}
where
\begin{eqnarray}
j(x<1)&=&\frac{4~{\rm
atanh}\sqrt{\frac{1-x}{1+x}}}{x^{2}\sqrt{1-x^{2}}}+\frac{2\ln\left(\frac{x}{2}\right)}{x^{2}}-\frac{1}{x^{2}-1}\nonumber\\
&+&\frac{2~{\rm atanh}\sqrt{\frac{1-x}{1+x}}}{\left(x^{2}-1\right)\sqrt{1-x^{2}}}\nonumber\\
j(x=1)&=&2\ln\left
(\frac{1}{2}\right )+\frac{5}{3}\\
j(x>1)&=&\frac{4{\rm
atan}\sqrt{\frac{x-1}{1+x}}}{x^{2}\sqrt{x^{2}-1}}+\frac{2\ln\left(\frac{x}{2}\right)}{x^{2}}-\frac{1}{x^{2}-1}\nonumber\\
&+&\frac{2~{\rm atan}\sqrt{\frac{x-1}{1+x}}}{(x^{2}-1)^{\frac{3}{2}}}\nonumber\;,
\end{eqnarray}
and the reduced shear $g=\frac{\gamma}{1-\kappa}$.

Throughout this section our ``standard" halo lens is at $z=0.2$, with 
parameters $c=6.0$ and $r_{200}=1.75~{\rm Mpc}$. The faint background galaxy 
population is at $z=1.0$. This corresponds to a rich galaxy cluster, with 
virial mass $M_{200}\sim 10^{15}M_{\odot}$. To determine the Einstein radius 
corresponding to particular values of $r_{200}$ and $c$ considered, 
$\bar\kappa(\theta_{\E})=1$ must be solved numerically; for the
standard parameters above, $\theta_{\E}\sim 0\arcminf 194$.

\begin{figure*}
\resizebox{12cm}{!}{\includegraphics{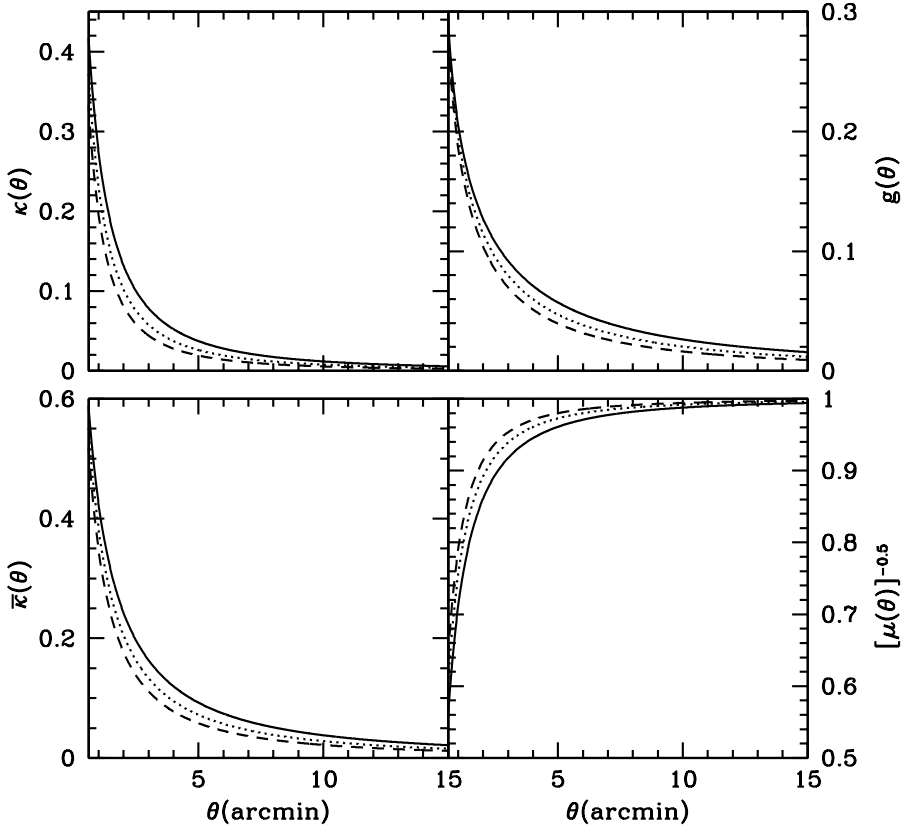}}
\hfill
\parbox[b]{55mm}{
\caption{In the four panels the convergence $\kappa$, mean convergence
$\bar\kappa$, reduced shear $g$ and magnification signal $\mu^{-0.5}$
are plotted as a function of radius $\theta$, for models with the
same $\theta_{\E}=0\arcminf 194$. The solid line represents a 
$c=5.0, r_{200}=1.967~{\rm Mpc}$ model, the short dashed line corresponds to
our standard $c=6.0, r_{200}=1.75~{\rm Mpc}$ model, and the long dashed line to
a $c=7.0, r_{200}=1.589~{\rm Mpc}$ model.
 }
\label{NFW3}}
\end{figure*}

\begin{figure}
\resizebox{8cm}{!}{\includegraphics{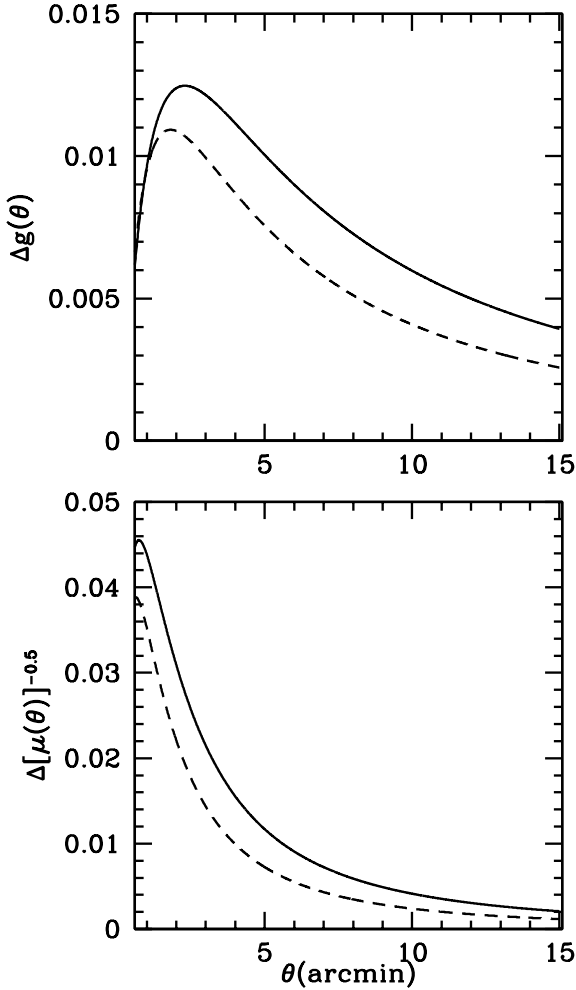}}
\hfill
\parbox[b]{88mm}{
\caption{This figure shows $\Delta g$ (upper panel) and
$\Delta\mu^{-0.5}$ (lower panel) as a function of
position in the aperture, $\theta$, for models with the same 
$\theta_{\E}=0\arcminf 194$. We compare our standard model, 
$c=6.0, r_{200}=1.75~{\rm Mpc}$, to a $c=5.0, r_{200}=1.967~{\rm Mpc}$ model
(solid line) and to a $c=7.0, r_{200}=1.589~{\rm Mpc}$ model (long dashed line). 
}
\label{NFW4}}
\end{figure}

Later, we drop the assumption that $\theta_{\E}$ is known and allow
$r_{200}$ and $c$ to vary independently. If $\theta_{\E}$ {\it is}
known, then the possible values of $c$ and 
$r_{200}$ are restricted to a curve in the $c-r_{200}$ plane. How well 
do we expect the magnification and shear methods to distinguish
between different values of $r_{200}$ and $c$ in this special case? 
Consider our standard model (i) $c=6.0, r_{200}=1.75~{\rm Mpc}$ and two
other models (ii) $c=5.0$, $r_{200}=1.967~{\rm Mpc}$ and 
(iii) $c=7.0$, $r_{200}=1.589~{\rm Mpc}$ which have the same $\theta_{\E}$. 
Fig.\ref{NFW3} compares $\kappa$, $\bar\kappa$, $g$ and $\mu^{-0.5}$ for
these models, as a function of $\theta$ between $0\arcminf 6$ and
$15\arcminf 0$. To see the differences between
the shear and magnification signals from the models, in Fig.\ref{NFW4} we plot
the absolute value of the difference, $|\Delta g|$ and $|\Delta\mu^{-0.5}|$,  
comparing our standard model (i) with models (ii) and (iii). The 
differences between the shear and magnification signals for these
models are quite small. A quick estimate shows that $\sim 8000$
galaxies are required to measure $\Delta g\sim 0.01$ at a
signal-to-noise of 3, which is possible with large format ground based
cameras. 
 
\subsection{SIE profile}

Very few clusters are perfectly spherical; therefore it is interesting
to find out whether one can take a catalogue of lensed galaxies and 
detect an ellipticity, $f$, in the mass distribution. We address this
question in Section \ref{sies}. 

Kormann, Schneider \& Bartelmann (1994) discussed the SIE profile,
with axial ratio $0<f\le 1$. Let the polar coordinates in
the lens plane be ${\vc\theta} = (\theta{\rm cos}\phi,\theta{\rm sin}\phi)$. 
The equivalent angular Einstein radius for a lens with velocity dispersion $v$ is 
$\theta_{\rm E} = 4\pi\frac{v^{2}}{c^{2}}\frac{D_{\rm ds}}{D_{\rm s}}$
and distances $x\equiv\vc\theta/\theta_{\rm E}$. 

The dimensionless surface mass density is given by
\begin{equation}
\kappa(x,\phi)=\frac{\sqrt{f}}{2b};~~~~~b=\sqrt{x_{1}^{2}+f^{2}x_{2}^{2}}\;,
\label{pop}
\end{equation}
the magnification is
\begin{equation}
\mu(x,\phi)=\frac{1}{1-2\kappa(x,\phi)}\;,
\end{equation}
and the components of the shear are
\begin{equation}
\gamma_{1}=-\kappa~{\rm cos}(2\phi);~~~~~\gamma_{2}=-\kappa~{\rm sin}(2\phi)\;.
\label{gammy}
\end{equation}

In the case of individual galaxies, comparison of the luminous mass
distribution with strong lens models makes it clear that the misalignment
of the axes of the luminous and dark matter distributions is less than 
$\sim10^{\circ}$ (e.g. Keeton et al. 1998). However, since for
clusters this need not be the case, we can drop the assumption that
the orientation is known, and make this a parameter in the model.
To do so, we recast $b$ given in Eq.(\ref{pop}) in polar coordinates,
and introduce the position angle of the major axis, $\alpha$
\begin{equation}
b=x\left[\left(1+f^{2}\right)+\frac{1}{2}\left(1-f^{2}\right){\rm cos}(2(\phi-\alpha))\right]^{\frac{1}{2}}\;. 
\end{equation}
Note that the dependence of the shear components given in Eq.(\ref{gammy})
is still on $\phi$ since the phase of $\gamma$ is the same when the
lens is rotated (and the same as in the circularly symmetric case).

\section{Results from the likelihood analysis}

\subsection{NFW profile: Numerical simulations to compare the likelihood analysis with $\chi^{2}$ statistics}

To check the validity of the analytic results, SKE presented a 
detailed comparison of the best-fit parameters obtained by applying 
the likelihood analysis to synthetic data sets, 
with the ensemble averaged log-likelihood contours. The introduction
of a new lens model will not affect the applicability of our
analytical techniques, but for completeness we demonstrate for one
case, that the analytical expressions are consistent with the Monte Carlo 
treatment, and that the errors on the recovered parameters follow a
$\chi^{2}$ distribution. An NFW lens model with true parameters
$\pi_{\t}: c=6.0,$ $r_{200}=1.75~{\rm Mpc}$ at $z_{\d}=0.2$ was used to generate
the catalogues of lensed galaxies and to derive the ensemble averaged 
log-likelihood contours.

\begin{figure*}
\resizebox{12cm}{!}{\includegraphics{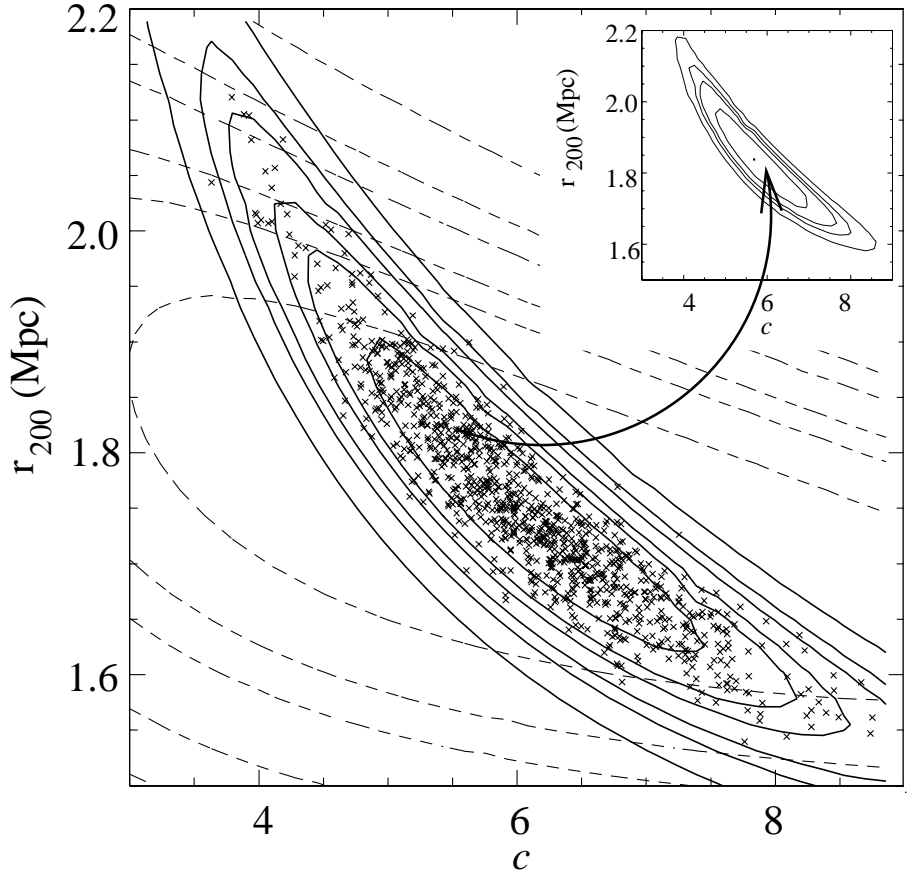}}
\hfill
\parbox[b]{55mm}{
\caption{In this figure, the input lens model has an NFW profile with
$\pi_{\t}$ of $c=6.0$
and r$_{200}=1.75~{\rm Mpc}$, and the inner and outer radii of the
annulus are $\theta_{\rm in}=0\arcminf 6$, $\theta_{\rm out}=4\arcminf
0$. The lens redshift is $z_{\d}=0.2$, and the other parameters are as 
described in the text. In the main panel, solid lines are contours of 
constant $\ave{\ell_{\gamma}}$ and dashed contours 
correspond to $\ave{\ell_{\mu}}$. Contours are drawn for 
$2\ave{\Delta\ell}=\{2.30, 4.61, 6.17, 9.21, 11.8, 18.4\}$, 
within which one expects that 68.3\%, 90\%, 95.4\%, 99\%,
99.73\% and 99.99\% respectively, of parameter estimates from
realisations will be enclosed. The crosses correspond to $\pi_{\rm max}$ 
recovered by applying the shear likelihood analysis to 
1000 simulated data sets. 
The insert panel corresponds to a single, randomly chosen, realisation 
$\pi_{\rm max}$ with corresponding $\ell(\pi_{\rm max})$. 
Contours of constant $\ell_{\gamma}$ are drawn for 
$2({\ell}(\pi)-{\ell}(\pi_{\rm max}))=\{2.30, 4.61, 6.17, 9.21\}$.}
\label{NFWealf}}
\end{figure*}

The main panel of Fig.\ref{NFWealf} shows the best-fit
parameters recovered by applying the shear likelihood analysis to 1000
synthetic catalogues, superimposed on the ensemble averaged
log-likelihood contours. Note that the scatter of the points in
the $c-r_{200}$ plane is consistent with the ensemble averaged
log-likelihood contours. The contours for the magnification method are 
wider than those for the shear method, although under the assumption
that the unlensed number density is accurately known, addition of the 
magnification information tightens the constraint on $r_{200}$ for
higher confidence levels. 

In order to verify that the errors on a single realisation are
comparable to those predicted by the ensemble average treatment,
individual catalogues with best-fit 
$\pi_{\rm max}$ and corresponding ${\ell}(\pi_{\rm max})$ can be
randomly selected and scrutinised. To illustrate this, the insert 
panel of Fig.\ref{NFWealf} shows contours of constant
$2({\ell}(\pi)-{\ell}(\pi_{\rm max}))$ for a random catalogue. 
As expected, we see that the confidence-levels for this individual
realisation are consistent with the ensemble averaged contours plotted
in the main panel. 

\begin{figure}
\resizebox{8.8cm}{!}{\includegraphics{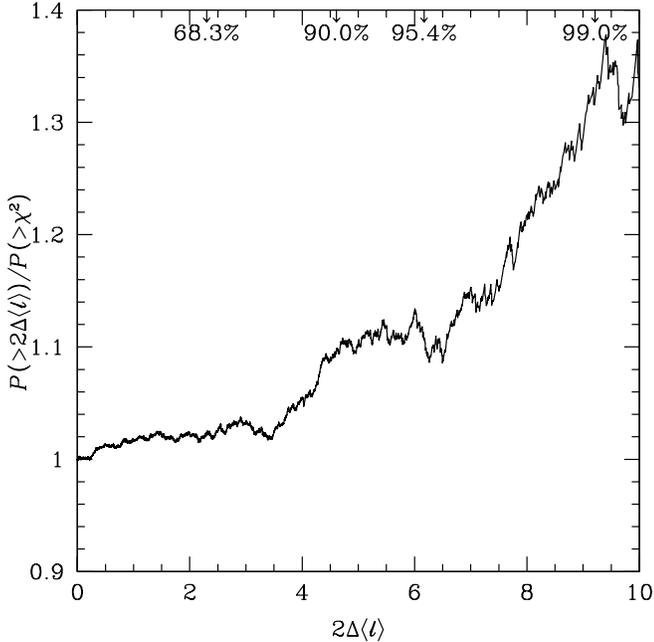}}
\hfill
\parbox[b]{88mm}{
\caption{
For 10000 simulated data sets, best-fit parameters were recovered
using the shear method. The ratio
of $P(>2\Delta\ave{\ell_{\gamma}})$ to that expected if
$2\Delta\ave{\ell_{\gamma}}$ followed a $\chi^{2}$ distribution is
shown as a function of $2\Delta\ave{\ell_{\gamma}}$. On the
top edge of the plot the 68.3\%-, 90\%-, 95.4\%- and 99\%-confidence
intervals are marked.
}
\label{NFWstat}}
\end{figure}

We now quantify the agreement between the
distribution of
 $2\Delta\ave{\ell} = \ave{\ell}(\pi) -\ave{\ell}(\pi_{\t})$ and that expected if $2\Delta\ave{\ell}$ followed
a perfect $\chi^{2}_{2}$ distribution. Ten thousand catalogues of lensed
galaxies were generated using the NFW lens model, with true
parameters $\pi_{\t}$ describing the cluster, and the log-likelihood
functions were minimised to obtain the best-fitting parameters
$\pi_{\rm max}$
for each realisation. For each $\pi_{\rm max}$ we then calculate
$2\Delta\ave{\ell}$ and derive the cumulative probability distribution
$P(>2\Delta\ave{\ell})$, which can be compared with the perfect
distribution $P(>\chi^{2}_{2})$. In Fig.\ref{NFWstat} the ratio of
$P(>2\Delta\ave{\ell})/P(>\chi^{2}_{2}$) is plotted against
$2\Delta\ave{\ell}$ for the shear method. The deviation from 
a $\chi^{2}_{2}$ distribution is small: it is less than $4\%$ until 
the 90\%-confidence interval, and even at the 95.4\%-confidence
interval there is only a $\sim 10\%$ deviation. These small deviations
from $\chi^{2}$ statistics can be attributed to the use of the 
analytic approximation (14) in obtaining 
$\ave{\ell_{\gamma}}$, whereas this approximation is not made during
the numerical simulations.

\subsection{Can we distinguish between NFW and power-law profiles?}

Two models commonly used to describe the radial dependence of the surface
mass density are the NFW and power-law profiles, a specific case of
which is the isothermal model. We briefly consider
whether the shear method enables a distinction to be made between these
profiles, and how much of a discrepancy there is between the best-fit
models of each family. 

The standard NFW lens was used to generate 
500 catalogues of lensed galaxies, with $n_{\gamma}=30~{\rm arcmin}^{-2}$. 
Best-fit parameters were recovered in an aperture with
inner radius $0\arcminf 6$ and outer radius $4\arcminf 0$, using the shear
method and assuming (i) an NFW lens (parameters $c,r_{200}$),
(ii) a lens with an isothermal slope, defined outside $\theta_{\E}$
[parameters $a$ (normalisation at $\theta_{\E}$) and $\theta_{\E}$]
and (iii) a power-law lens (parameters $a$ and slope $q$), with 
$\theta_{\E}$ assumed to be known from observations, and set equal to 
that of the 
true NFW model. We refer to quantities associated with recovery under the
true NFW model with the subscript ``true", and to those under other
models with the subscript ``false". 

Comparing $\ell_{\rm max, true}$ and $\ell_{\rm max, false}$ on a catalogue by 
catalogue basis shows that the NFW profile has a formally higher likelihood 
than the isothermal profile in 97.6\% of cases. In practice, for 
a single data set, a particular profile would be considered to better 
describe the lens if it had a likelihood in excess of $\sim 2\sigma$ (depending on the observational noise) of other models considered. 
Evaluation of
$2(\ell_{\rm max, true}-\ell_{\rm max, false})$ and comparison with the standard
confidence-levels for normal distributions with 2 degrees of 
freedom shows that for 88.3 (71.7, 58.4, 39.6)\% of the cases where the
NFW profile is the most likely model, the ability to distinguish it
from the isothermal profile is at greater than 
68.3 (90, 95.4, 99)\% confidence. In other words, for this case,
distinguishing the NFW and isothermal models at 2$\sigma$ confidence is possible in about 60\% of cases where the NFW model has a higher likelihood. These results are summarised in Table\ref{nfwisoconf}. In cases where the 
isothermal fit is formally better, none of these realisations reach 
68.3\% confidence. 

\begin{table}
\begin{tabular}{|l|l|}
\hline
$\Upsilon$(\%)&P$(>\Upsilon)$(\%)\\
\hline
68.3&88.3\\
90.0&71.7\\
95.4&58.4\\
99.0&39.6\\
\hline
\end{tabular}
\caption
{
The standard NFW lens was used to generate 500 catalogues of lensed galaxies, with $n_{\gamma}=30~{\rm arcmin}^{-2}$. Best-fit
parameters were recovered in an aperture with $\theta_{\rm
in}=0\arcminf 6$ and $\theta_{\rm out}=4\arcminf 0$. An NFW lens and
an isothermal lens (parameterised as described in the text) were
independently fit to the catalogues and their likelihoods compared.
Considering cases where the NFW profile has a formally higher
likelihood, and denoting the confidence level at which NFW and
isothermal models can be distinguished by $\Upsilon$, 
the table shows P$(>\Upsilon)$ as a function of $\Upsilon$.
\label{nfwisoconf}}
\end{table}

In the case of recovery with the more general power-law model, the 
best-fit NFW model has a 
higher likelihood in 67.8\% of realisations, with 23.6\% of these
realisations being above the 68.3\% confidence-level. For those 
realisations where the
power-law model has a formally higher likelihood, a smaller fraction
(10.6\%) are above the 68.3\% confidence-level, which is expected from
the form of the distribution of $2(\ell_{\rm max, true}-\ell_{\rm max, false})$.

\begin{figure}
\resizebox{8.8cm}{!}{\includegraphics{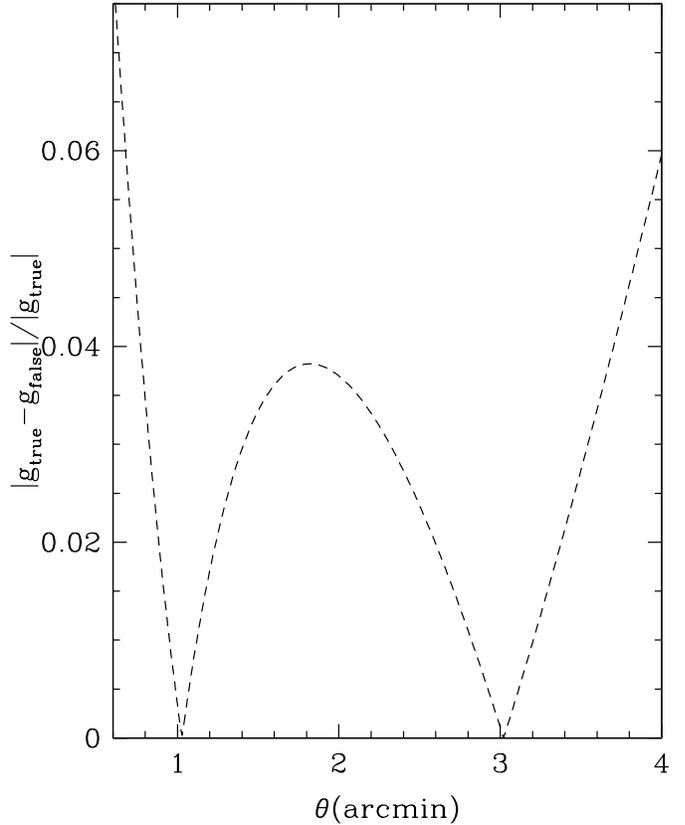}}
\hfill
\parbox[b]{88mm}{
\caption{
The standard NFW lens was used to generate 500 catalogues of lensed
galaxies. The best-fit parameters were recovered using the shear
method in an aperture with $\theta_{\rm in}=0\arcminf 6$,
$\theta_{\rm out}=4\arcminf 0$, and assuming (i) an NFW lens and 
(ii) a power-law lens. The figure shows $|g_{\rm true}-g_{\rm
false}|/|g_{\rm true}|$ as a function of $\theta$, distance from the
aperture centre. These values were obtained at discrete values of 
$\theta$ by taking the arithmetic mean of $g(\theta)$ for the 500 
best-fit models.}
\label{nfwpow}}
\end{figure}

Just how different are the profiles corresponding to the best-fit parameters? 
To get an idea of how the statistics translate into observable
differences, consider models obtained by taking the arithmetic mean of 
$\pi_{\rm max}$ at discrete distances from the lens centre,
$\theta$ for the NFW parameters (the mean recovered values of
$c$ and of $r_{200}$) and 
for the power-law models (the mean recovered values of $a$ and of $q$). 
The value of $|g(\vc\theta)_{\rm true}-g(\vc\theta)_{\rm false}|/|g(\vc\theta)_{\rm true}|$ is 
shown in Fig.\ref{nfwpow} as a function of $\theta$. At small radii the percentage difference is fairly large (but in this
region it is observationally difficult to make measurements), dropping to
zero at intermediate radii before increasing again to around 4\%, 
dropping again to zero before slowly increasing at larger radii. 

For illustration, $\theta_{\rm out}$ was increased to $15\arcminf 0$ and
100 catalogues were generated using the NFW lens, and again the best-fit
parameters were recovered using the two-parameter power-law model during
the analysis. Comparison of the best-fit models shows that the 
NFW profile has a higher likelihood in 96\% of cases; in 95 (82)\% of these
cases, the difference from the best power-law model is at greater 
than 68.3 (95.4)\% confidence. When $\theta_{\rm out}$ is increased,
note the marked increase in the percentage of cases where the true model has a
higher likelihood, and at a higher confidence level. 


In the future, space-based telecopes with a wide field-of-view
will be available for studies of lensing clusters. To mimic the action
of such a telescope, catalogue generation with the NFW profile and 
recovery with the NFW and power-law profiles was performed with 
$n_{\gamma}=150~{\rm arcmin}^{-2}$, with $\theta_{\rm out}=15\arcminf 0$, 
giving $\approx\sqrt 5$ increase in
the signal-to-noise beyond that of the lower density case. With this 
extreme background density and data field size, {\em all} of the NFW best-fit 
models have higher likelihoods than the corresponding best-fit
power-law models, at 99.99\% significance difference. 
Another possible means to effectively increase 
$n_{\gamma}$ is by ``stacking" large enough samples of ground
based wide-field images of clusters. An analagous process has been
undertaken for galaxy groups [see Hoekstra et al. 1999].
   

\subsection{Singular ellipsoidal profile\label{sies}}

For this profile, the primary aim was to see how well the axial ratio
and position angle of the SIE mass distribution could be recovered. 
The SIS is a special case of the SIE model ($f=1.0$), so we can also
see how likely it is that an SIE will be misidentified as an SIS.

\begin{figure*}
\resizebox{12cm}{!}{\includegraphics{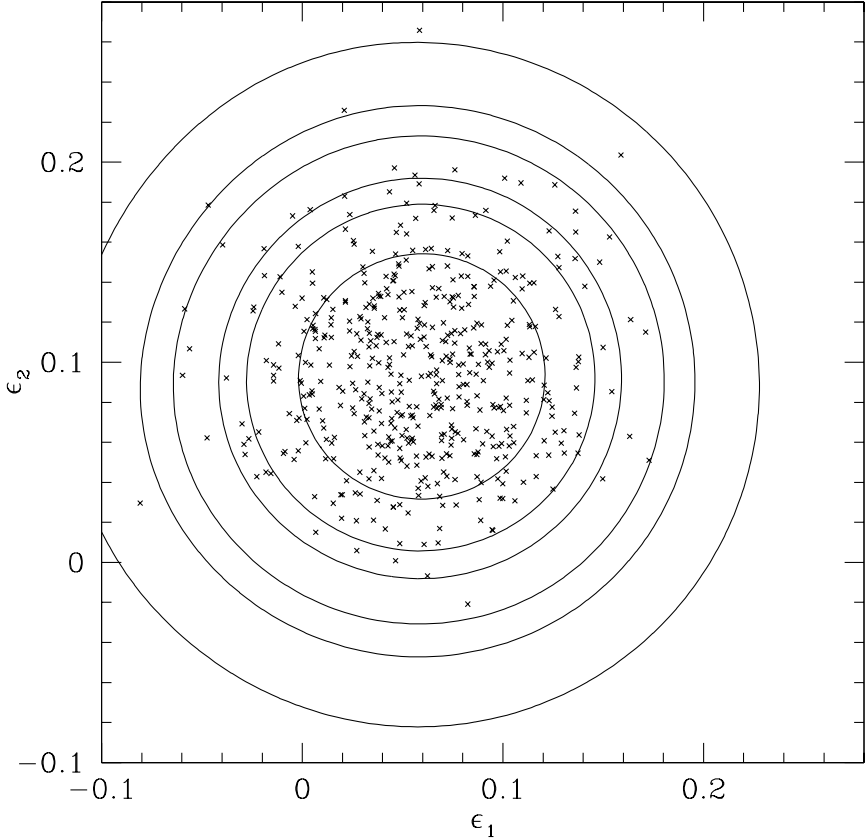}}
\hfill
\parbox[b]{55mm}{
\caption{
An SIE lens at $z_{\d}=0.2$ with $\theta_{\rm E}=0\arcminf 5$ and $\pi_{\t}$: $f=0.8$,
$\alpha=0.5~{\rm radians}$ was used to generate 500
catalogues of lensed galaxies. The best-fit parameters were recovered
using the shear method in an aperture with $\theta_{\rm in}=1\arcminf 0, \theta_{\rm out}=3\arcminf 0$. The panel shows $\epsilon_{1}$ and
$\epsilon_{2}$ (the components of the complex ellipticity of the
cluster) for the $\pi_{\rm max}$ (crosses). Contours of constant
$\ave{\ell_{\gamma}}$ for $2\Delta\ell=\{2.3, 4.61, 6.17, 9.21, 11.8,
18.4\}$
are indicated by the solid lines. 
}
\label{ellipso}}
\end{figure*}

We considered the situation where the orientation of the SIE cluster
is not well determined -- i.e. where the free parameters are $f$ and 
the position angle $\alpha$. The values of $\pi_{\rm max}$ were recovered
for 500 realisations, when $\alpha$ was set to 0.5~radians 
($\approx 28.6^{\circ}$). A convenient representation of a cluster's 
ellipticity is to write it in complex form as
\begin{equation}
\epsilon=\frac{1-f}{1+f}{\rm e}^{2{\rm i}\alpha}\equiv F{\rm e}^{2{\rm i}\alpha}.
\end{equation}
Then the log-likelihood minimisation can be performed in the
$\epsilon_{1}-\epsilon_{2}$ plane and in Fig.\ref{ellipso} we mark the best-fit parameters $\pi_{\rm max}$
($\eps_{1}=F{\rm cos}(2\alpha)$ and $\eps_{2}=F{\rm sin}(2\alpha)$) with
crosses (in this representation, $\pi_{\t} \equiv (0.06, 0.0935)$). 
Note that $\alpha_{\rm max}=0.5~{\rm tan}^{-1}(\epsilon_{2}/\epsilon_{1})$
and that the distance from the origin is $F$.

The ensemble averaged log-likelihood function can be determined for
the SIE by performing a 2-D numerical integration, either over the 
$f-\alpha$ plane or over the $\epsilon_{1}-\epsilon_{2}$ plane. Contours of
constant $2\ave{\ell_{\gamma}}$ calculated in the
$\epsilon_{1}-\epsilon_{2}$ plane are superimposed on
Fig.\ref{ellipso}. Again, as in the
case of circularly symmetric models, the spatial distribution of
the realisations agrees well with the likelihood contours. The value
of $f$ is well constrained - for instance, SIS models with $f=1.0$ lie 
outside the 95.4\% confidence-level. As the value of $\alpha_{\rm
true}$ is changed,
then the locus of the centre of the ensemble averaged contours is a
circle centred on the origin in the $\epsilon_{1}-\epsilon_{2}$ plane, 
and the areas of the confidence regions remain unchanged.  

When $f_{\t}=1.0$ (i.e. the SIS case) $\alpha$ is
unconstrained since the profile is circular; as $f_{\t}$ is decreased, 
the ability to constrain $\alpha$ increases.
  
\section{Redshift distribution}

Throughout this Section, the catalogue generation and best-fit model recovery 
were performed using a cluster described by an NFW profile, $\pi_{\rm t}$: $c=4.0$, $r_{200}=1.0~{\rm Mpc}$, at $z_{\d}=0.5$. The number density of
sources was taken to be $n_{\gamma}=40~{\rm arcmin}^{-2}$. Five hundred 
catalogues
of lensed galaxies were generated in an aperture with 
$\theta_{\rm in}=0\arcminf 2$ and $\theta_{\rm out}=6\arcminf 0$, 
following the prescription given in Section 4. The best-fit parameters 
were recovered independently under each redshift knowledge scenario
outlined in Section \ref{redinc}. 

When photometric redshifts ($z_{\rm ph}$) were assumed to be
available, these were assigned by
adding a Gaussian random deviate of dispersion $\sigma_{z}=0.1$ to the
true redshift $z_{t}$. During the analysis, the photometric estimates
were inserted in place of the true redshifts. As mentioned previously, 
their dispersion can be thought of as a modification of
$\sigma_{\eps}$, giving an effective dispersion $\sigma_{\rm eff}$. 
From our analytic estimate, we
would expect that the error on parameters recovered would not differ 
much from the ideal case. This has also been indicated by Bartelmann
\& Schneider (2000).

In the case where the galaxies were placed on a sheet, the value of
$\ave{w}$ was determined for the lens at $z_{\d}=0.5$. For $p(z){\d}z$ 
in question, $\ave{w}=0.35$, which corresponds to $z_{\rm sheet}=0.82$. 
The values $\ave{w}^{2}=0.125$ and $\ave{w^{2}}=0.185$ imply that 
$\ave{g_{t}}/\ave{g_{\rm sheet}}\approx 1+0.48\kappa$; this means 
that the cluster mass is overestimated with the sheet approximation, 
since it must be more massive to produce the same lensing signal.
For typical values of $\kappa\sim 0.1$, the correction is fairly
small.

When the form of the redshift distribution is known, but not the
redshifts of individual galaxies, Eq.(\ref{redinfo}) was discretised 
and integrated numerically. 

\begin{figure*}
\resizebox{12cm}{!}{\includegraphics{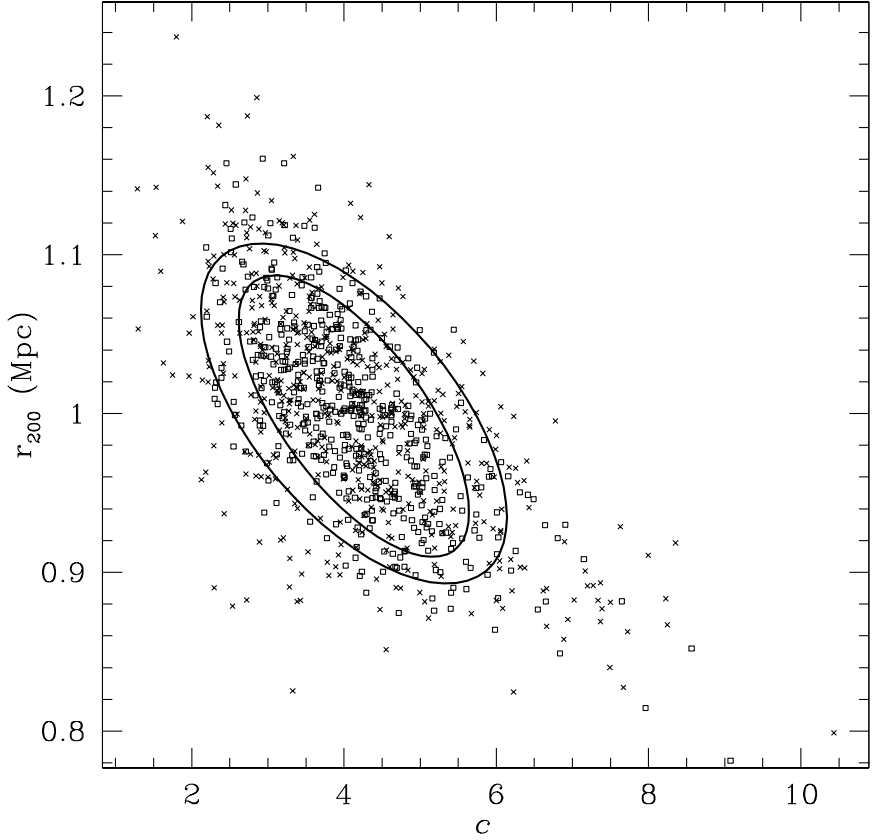}}
\hfill
\parbox[b]{55mm}{
\caption{
An NFW lens at $z_{\d}=0.5$, $\pi_{t}$: $c=4.0$, $r_{200}=1.0~{\rm Mpc}$, was
used to generate 500
catalogues of lensed galaxies, where $p(z){\rm d}z$ for the source galaxies is
as described in the text. Best-fit parameters were recovered
using the shear method in an aperture with $\theta_{in}=0\arcminf 2$, 
$\theta_{\rm out}=6\arcminf 0$. The squares show parameters recovered
when the true redshifts are known individually and the crosses show
recovery when the source galaxies are placed at $z_{\rm sheet}$. 
The 68.3\% confidence ellipses were determined from the eigenvectors
and eigenvalues of the error matrices, and are shown for the true redshift
(inner ellipse) and sheet case (outer ellipse).}
\label{reds}}
\end{figure*}

Fig.\ref{reds} shows the recovered $\pi_{\rm max}$ for the ideal
case when all redshifts are known (squares) and for the case where
the sources are at $z_{\rm sheet}$ (crosses), to give an indication of 
the scatter. Error matrices were obtained from the values 
of $\pi_{\rm max}$ for each case. Denoting $(c-c_{\t})$ by ${\cal G}_{1}$ and
$(r_{200}-r_{200_\t})$ by ${\cal G}_{2}$, then the elements 
are ${\cal G}_{ij} \equiv \ave{{\cal G}_{i}{\cal G}_{j}}$. Then,
error ellipses were obtained from the eigenvalues and eigenvectors of
these matrices. The inner ellipse corresponds to the true redshift
case, and the outer ellipse to the sheet case. 

The relative error matrices were 
derived from the values of $\pi_{\rm max}$ for each of the redshift
assumptions. Denoting $(c-c_{\t})/c_{\t}$ by ${\cal E}_{1}$ and
$(r_{200}-r_{200_\t})/r_{200_\t}$ by ${\cal E}_{2}$, then the elements 
are ${\cal E}_{ij} \equiv \ave{{\cal E}_{i}{\cal E}_{j}}$. 
The elements of the relative error matrices are shown in Table\ref{E}.
 
Naturally, when all the source redshifts are known, the dispersion 
of $\pi_{\rm max}$ 
is smallest. If photometric redshifts are available, then the
dispersion is only marginally greater $(\sim 1.5\%)$, which we expected from our 
estimate. The dispersion becomes greater when only the 
distribution is known $(\sim 21\%)$, or greater still when the 
sources are assumed to be on a sheet $(\sim 60\%)$.
The advantage of having redshift information becomes more paramount as
the redshift of the lens is increased -- recall from before that if a
cluster is at a low redshift, then the redshift distribution is 
fairly unimportant. Also, if $\kappa$ is large this information is also more
important since the errors incurred by making a sheet approximation
become large.


\begin{table}
\begin{tabular}{|l|l|l|l|}
\hline
Redshift Scenario&${\cal E}_{11}$&${\cal E}_{12}$(Mpc)&${\cal E}_{22}$(Mpc$^{2}$)\\
\hline
True redshifts&0.063&-0.0101&0.00345\\
Photometric redshifts&0.063&-0.0103&0.00358\\
Redshift distribution&0.069&-0.014&0.007\\
Sheet&0.11&-0.014&0.005\\ 
\hline
\end{tabular}
\caption{
An NFW cluster at $z_{\d}=0.5$ and with $\pi_{t}$: $c=4.0$, $r_{200}=1.0~{\rm Mpc}$ was used
to generate 500 catalogues of lensed galaxies. Best-fit parameters were
recovered under different assumptions of redshift knowledge, indicated
in the left-hand column. The elements of the relative error matrices (see text
for details) are given in the right-hand columns.}
\label{E}
\end{table}

We can compare these results with the analytic work presented in 
Bartelmann \& Schneider (2000) who focused on the
dispersion of shear estimates in the cases where all redshifts are 
known, and where only the distribution $p(z){\d}z$ is known. Although 
many of their input parameters are different, a qualitative comparison can
be made. For instance, at $z_{\d}=0.5$ they find an improvement of
$30\%$ in the accuracy of the shear estimate when redshifts are
known, as opposed to when only $p(z){\d}z$ is known - we obtain a
similar result: $\sim 21\%$ 
improvement in the dispersion of recovered $\pi_{\rm max}$ between
these two cases. 

\section{The influence of substructure}

In the preceeding sections we considered smooth parametric models of
clusters, for analytical and numerical simplicity. The full treatment
of substructure requires studying how its 
spatial distribution and power spectrum interplay with the smooth
cluster profile and the aperture within which the observations are 
made. This is beyond the scope of this work and will be investigated in 
a forthcoming paper. Here we illustrate the implications of substructure on
the analysis, by constructing a
non-smooth toy model consisting of a smooth NFW 
cluster ($z_{\d}=0.2$, $c=6.0$ and $r_{200}=1.75~{\rm Mpc}$), with
surface mass density $\kappa_{\rm smooth}(r)$, which is modified by the
addition of smaller scale sub-profiles, denoted by $\kappa_{\rm sub}(r)$. In order to
preserve the overall mass of the cluster, $\kappa_{\rm sub}(r)$ has to be
chosen so that $2\pi\int_{0}^{r_{\rm sub}}r\kappa_{\rm
sub}(r){\d}r=0$, where $\kappa_{\rm sub}$ is defined for $r<r_{\rm sub}$. A suitable functional form is
%
\begin{eqnarray}
\kappa_{\rm sub}(r)&=&\kappa_{\rm sub_{c}}\frac{16}{\pi}\left(0.25-\left(\frac{r}{r_{\rm sub}}\right)^{2}\right)\nonumber\\
&\times&
\left(1-\left(\frac{r}{r_{\rm sub}}\right)^{2}\right)^{2}, 
\end{eqnarray}

where $(4/\pi)\kappa_{\rm sub_{c}}$ is the central surface mass
density. Below, we take $r_{\rm sub}=1\arcminf 0$ and amplitude
scales $\kappa_{\rm sub_{c}}= 0.01, 0.025$ and 0.05. The magnitude of
the corresponding shear field is
\begin{eqnarray}
|\gamma_{\rm sub}|(r)&=&\kappa_{\rm
 sub_{c}}\frac{4}{\pi}\biggl(3\left(\frac{r}{r_{\rm
 sub}}\right)^{2}-6\left(\frac{r}{r_{\rm
 sub}}\right)^{4}\nonumber\\
&+&3\left(\frac{r}{r_{\rm sub}}\right)^{6}\biggr),
\end{eqnarray}
and the phase is determined from the position angle relative to the
centre of the sub-profile. The final surface mass
density is simply a linear superposition of the smooth and $n$ sub-profiles, 
\begin{equation}
\kappa_{\rm tot}(r)= \kappa_{\rm smooth}(r)+ \sum_{n}\kappa_{\rm sub}(r),
\end{equation}
and similarly for the shear field $\gamma$,
\begin{equation}
\gamma_{\rm tot}(r)=\gamma_{\rm smooth}(r)+ \sum_{n}\gamma_{\rm sub}(r).
\end{equation}
Finally, the reduced shear is
\begin{equation}
g_{\rm tot}(r)=\frac{\gamma_{\rm tot}(r)}{1-\kappa_{\rm tot}}.
\end{equation}

\begin{figure}
\resizebox{8.8cm}{!}{\includegraphics{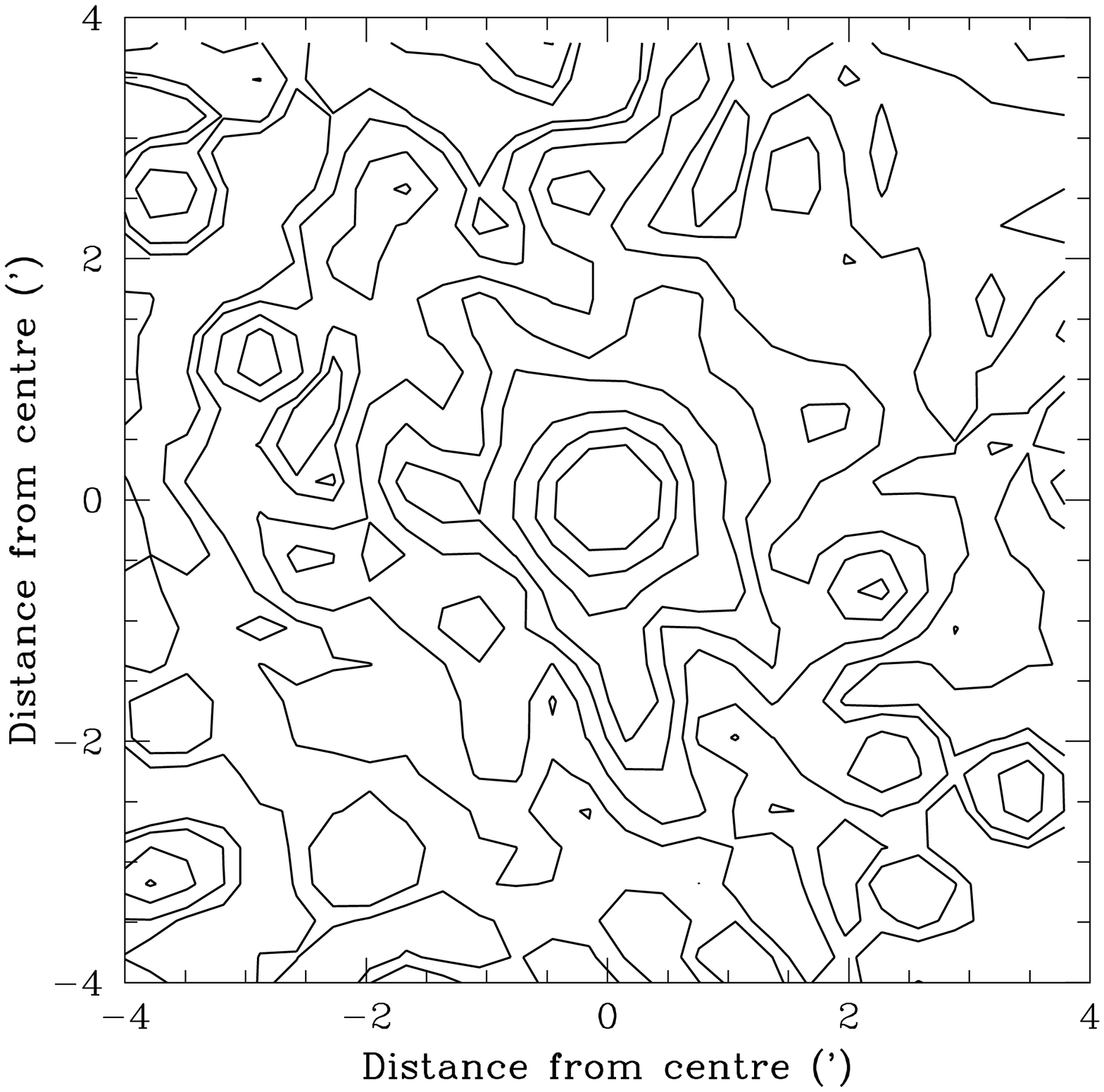}}
\hfill
\parbox[b]{88mm}{
\caption{This figure shows $\kappa_{\rm tot}$ over an 8$\arcminf$0
field, resulting when 
500 sub-profiles of amplitude $\kappa_{\rm sub_{c}}= 0.05$ (see text
for further details) are added to an NFW cluster 
($z_{\d}=0.2$, $c=6.0$ and $r_{200}=1.75~{\rm Mpc}$). The contours
shown correspond to $\kappa_{\rm tot}=0.025, 0.05, 0.1, 0.15, 0.2,
0.3, 0.4, 0.5$. }
\label{kappa}}
\end{figure}

Catalogues of lensed galaxies were generated using
the prescription above, with 500 equal amplitude sub-profiles 
distributed between $0\arcminf 0$ and $15\arcminf 0$ from the cluster
centre, their number density decreasing with radius. 
The resulting $\kappa_{\rm tot}$ inside a $4\arcminf 0$
radius is shown in Fig.\ref{kappa}, for $\kappa_{\rm sub_{c}}= 0.05$. 
The shear likelihood analysis was performed to obtain the best-fit
smooth model consistent with the data. 

With substructure the advantages of using data close to the centre of
the lens, are counteracted by the disadvantage of the increased
deviation from the smooth value of $g_{\rm tangential}(r)$. Here we
restrict our consideration to one of our standard apertures with 
$\theta_{\rm in}=0\arcminf 6$ and $\theta_{\rm out}=4\arcminf 0$.

First of all, in generating each lensed catalogue we allowed both the
spatial distribution of the substructure, and the spatial and
ellipticity distribution of
the background galaxies to vary. In the second instance, the 
distribution of background galaxies was kept fixed, so that the
contribution of randomising the substructure to the results
could be disentangled. We recovered $\pi_{\rm max}$ for the smooth 
input profile, and for input profiles with 
$\kappa_{\rm sub_{c}}= 0.01, 0.025$ and 0.05.

\begin{figure*}
\resizebox{12cm}{!}{\includegraphics{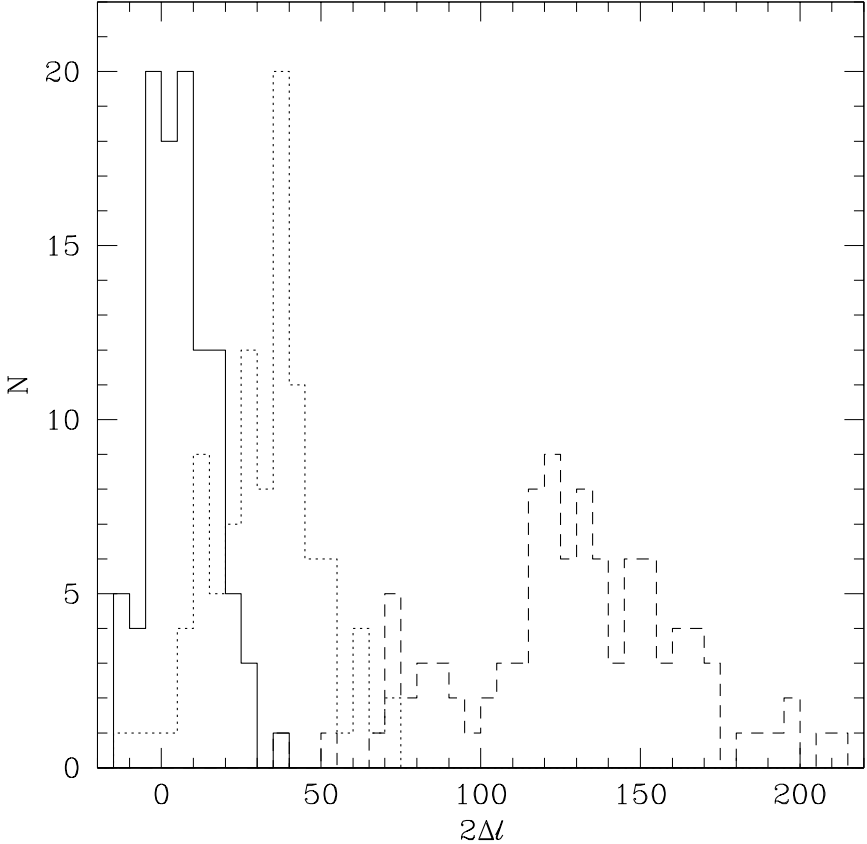}}
\hfill
\parbox[b]{55mm}{
\caption{
Various amplitudes of sub-profiles were added to 
an NFW cluster at $z_{\d}=0.2$, $\pi_{t}$: $c=6.0$,
$r_{200}=1.75~{\rm Mpc}$, which was then used to generate 100 
catalogues of lensed galaxies and the best-fit smooth model was recovered.
This figure shows the distribution of $2\Delta\ell$,
calculated for $\kappa_{\rm sub_{c}}= 0.01$ (solid line), 0.025 (short
dashed line) and 0.05 (long dashed line).}
\label{likely}}
\end{figure*}

The relative error matrices were derived from the values of $\pi_{\rm max}$ for
each of the smooth and substructure cases. Denoting $(c-c_{\t})/c_\t$ by 
${\cal E}_{1}$ and $(r_{200}-r_{200_\t})/r_{200_\t}$ by ${\cal E}_{2}$, 
then the elements of the matrices are 
${\cal E}_{ij} \equiv \ave{{\cal E}_{i}{\cal E}_{j}}$; these are given
in Table\ref{suberr}. 
One can also compare the log-likelihoods of the
$\pi_{\rm max}$ for each of the catalogues generated with substructure, and
the catalogue generated from the smooth profile (both analysed 
assuming a smooth profile, as mentioned above). Fig.\ref{likely} shows 
histograms of $2\Delta\ell$ for $\kappa_{\rm sub_{c}}= 0.01$ (solid
line), 0.025 (short dashed line) and 0.05 (long dashed line). 
In the case where the amplitude $\kappa_{\rm sub_{c}}= 0.01$, the
associated error matrix is statistically indistinguishable from that
of the smooth profile. As the amplitude increases, the error on the
recovered parameters increases as does $2\Delta\ell$.

\begin{table}
\begin{tabular}{|l|l|l|l|}
\hline
Substructure Scenario&${\cal E}_{11}$&${\cal E}_{12}$(Mpc)&${\cal E}_{22}$(Mpc$^{2}$)\\
\hline
{\em A}&0.018&-0.0069&0.0030\\
{\em B}&0.021&-0.0070&0.0028\\
{\em C}&0.033&-0.0089&0.0029\\
{\em D}&0.083&-0.0173&0.0042\\
{\em E}&0.055&-0.0083&0.0018\\
\hline
\end{tabular}
\caption{
An NFW cluster at $z_{\d}=0.2$, $\pi_{t}$: $c=6.0$,
$r_{200}=1.75~{\rm Mpc}$ and with various degrees of substructure 
was used to generate 100 catalogues of lensed galaxies. 
The scenarios indicated in the left-hand column are as follows: {\em A}
correponds to the smooth profile, and the random factor of 
each of the realisations is the spatial distribution and intrinsic
ellipticities of the background galaxies. For scenarios {\em B, C} and
{\em D}, 500 sub-profiles with $\kappa_{\rm sub_{c}}= 0.01, 0.025$ and 0.05
respectively were added to the
smooth profile and for each realisation the background galaxies and
the spatial distribution of the substructure were random. Finally, for {\em E} the background galaxy
distribution was kept fixed and only the spatial distribution of the 
substructure was randomised, with $\kappa_{\rm sub_{c}}=0.05$. The elements of the relative error
matrices (see text for details) are given in the right-hand columns.}
\label{suberr}
\end{table}

\section{Discussion and conclusions}

Fitting parameterised models to clusters is essential when a
statistical comparison is to be made between them. This is becoming
especially important with the explosion in the use of wide field
imagers to study cluster samples. The aims are to ascertain the
best method to constrain these models, and to be able to predict the
uncertainties in parameters obtained given observations of 
different depths, data field size and level of knowledge of the 
redshift distribution of the galaxy population. We find that
likelihood and ensemble averaged likelihood techniques are an
excellent means to achieve these goals. 

In this paper we compared the accuracy with which parameters can be
fit to lower redshift clusters described by NFW profiles (parameters $c$
and $r_{200}$), using the shear and magnification information. We find that\\ 
\noindent
(i) for the case considered, the shear method is superior, although there
is a marked degeneracy in the determination of the concentration
parameter, $c$.\\\noindent 
(ii) The analytic treatment is consistent with the results of Monte
Carlo simulations, and that the assumption of $\chi^{2}$-statistics is
a good approximation.

To investigate whether we can distinguish between NFW and power-law
models, catalogues of lensed galaxies were generated using an NFW
model, and parameters recovered under an NFW and power-law model
independently:\\\noindent 
(iii) The ability to distinguish between profiles depends
strongly on the size of the field-of-view used in the lensing
analysis, and on the number density of sources. For example, 
for the model considered, 
when $\theta_{\rm out}$ is increased from $4\arcminf 0$ to 
$15\arcminf 0$ there is a $\sim 40\%$ improvement in the ability to
distinguish between 2 parameter NFW and power-law models.

In practice, other factors that would have to be taken into account
include how close to the centre of the cluster it is possible to take
useful data, and other sources of noise in the observations. 
The number density of galaxies available for the analysis will depend
on the limiting magnitude of the observations, and the seeing 
conditions. As to the issue of the PSF anisotropy, which hampers the 
accurate measurement of galaxy shapes, detailed simulations using 
realistic PSF profiles have been undertaken by Erben et al. (2000); 
these show that gravitational shear can be recovered with an error of
10-15\%. Similar work is presented by Bacon et al. (2000). Gray et al. (2000) recently applied the maximum
likelihood magnification
method to infrared CIRSI observations of Abell 2219,
fitting single parameter SIS ($\theta_{\E}$) and NFW ($r_{\s}$)
profiles. However, with the noise in their observations it was not
possible to differentiate between the profiles. 

The position angle of the luminous and dark matter distributions may 
be significantly misaligned. By considering the SIE profile, we have shown\\\noindent
(iv) at what confidence the complex ellipticity of a cluster can 
be recovered,\\\noindent
(v) In the case of this non-circulary
symmetric profile, we have also demonstrated that our numerical
simulations are consistent with the ensemble averaged log-likelihood 
confidence-intervals.
 
We also examined the dispersion in recovered parameters for an NFW
cluster at a higher redshift ($z_{\rm d}=0.5$), assuming 
several possibilities for our knowledge of the redshift distribution of source 
galaxies. Since we are moving into an
era when photometric redshifts with $\sigma_{z}\approx 0.1$ are
becoming available for large samples of galaxies, a striking
conclusion is that\\\noindent
(vi) the fractional gain in the accuracy of parameter 
estimates is only $\sim 1.5\%$ more when exact redshifts are known
rather than photometric estimates.\\\noindent
(vii) If only $p(z){\rm d}z$ is known, then the fractional gain in having
spectroscopic redshifts is $\sim 21\%$.\\\noindent
(viii) The most common assumption in weak 
lensing analyses for the source redshift distribution - that they lie 
on a sheet whose redshift is determined by $\ave{w}$ - gives
parameters with the largest errors, and having spectroscopic or
photometric redshifts would decrease their dispersion by $\sim 60\%$. 
If colour or redshift information is available which enables 
some fraction of foreground or cluster galaxies to be identified and
excluded from the analysis, then the dispersion in parameter estimates becomes lower. 

Throughout for simplicity we assumed an EdS cosmology; 
qualitatively, our conclusions are unaffected by this and 
quantitatively the discrepancy is small. The cosmological weighting
function $w(z)$ is readily evaluated for any cosmological model, and 
the difference in an $\Omega_{0}=0.3, \Omega_{\Lambda}=0.7$ cosmology 
is less than 10\% even when the lens redshift $z_{\d}=0.8$ [see
Bartelmann \& Schneider (2000)]. In any case, changing the
cosmological model is equivalent to applying a small change to the
anyway uncertain redshift distribution.  

Finally, we qualitatively considered how fitting parameterised models 
to lensing data is influenced by the presence of substructure. Our 
preliminary work will be developed in a forthcoming paper. In the 
framework of our basic model, we find that\\\noindent
(ix)increasing the amplitude of the substructure increases the
dispersion of the recovered parameters.

\begin{acknowledgements}
We would like to thank Thomas Erben, Matthias Bartelmann, Douglas
Clowe, Marco Lombardi and Hojun Mo for very interesting discussions,
and Matthias for carefully reading the manuscript. We are grateful 
to the referee for very helpful comments which have helped to improve the
paper. This work was supported by the TMR Network
``Gravitational Lensing: New Constraints on Cosmology and the
Distribution of Dark Matter'' of the EC under contract
No. ERBFMRX-CT97-0172.
\end{acknowledgements}
  
\def\ref#1{\bibitem[1998]{}#1}

\end{document}